\documentclass[useAMS,usenatbib]{mnras}
\usepackage[fleqn]{amsmath}
\usepackage{graphicx,txfonts,fleqn,enumerate,color}

\newcommand*{\Msun}      {\text{M}_\odot} 
\newcommand*{\phm}       {\phantom{-}}
\newcommand*{\kms}		 {\mbox{km\,s$^{-1}$}}
\newcommand*{\kpc}		 {\mbox{kpc}}
\renewcommand*{\vec}[1]  {\boldsymbol{#1}}

\newcommand*{\sub}[2]    {{#1}_{\mathrm{#2}}}
\renewcommand*{\d}       {\mathrm{d}}

\newcommand*{\diff}[2]   {\frac{\mathrm{d} #1}{\mathrm{d} #2}}
\newcommand*{\tdiff}[2]  {\frac{\mathrm{d}^2{#1}}{\mathrm{d}{#2}^2}} 
\newcommand*{\pdiff}[2]  {\frac{\partial #1}{\partial #2}}

\newcommand*{\ptdiff}[3] {\frac{\partial^2\!#1}{\partial #2\,\partial #3}}
\newcommand*{\ptdiffx}[2]{\frac{\partial^2\!#1}{\partial#2^2}}

\title[Tidal ribbons]{Tidal ribbons}

\author[Dehnen \& Hasanuddin]{
Walter Dehnen$^{1}$\thanks{E-mail: wd11@leicester.ac.uk, h2@leicester.ac.uk}
and
Hasanuddin$^{1,2\star}$
\\
$^{1}$Department of Physics and Astronomy, University of Leicester, University Road, Leicester LE1 7RH, UK\\
$^{2}$Department of Physics, Faculty of Mathematics and Natural Sciences, Tanjungpura University, Prof. H. Hadari Nawawi St., Pontianak 78124, Indonesia
}

\date{Accepted XXX. Received YYY; in original form ZZZ}

\pubyear{2018}

\begin{document}
\label{firstpage}
\pagerange{\pageref{firstpage}--\pageref{lastpage}}
\maketitle

\begin{abstract}
Tidal debris from Galactic satellites generally forms one-dimensional elongated  streams, since nearby Galactic orbits have almost identical frequency ratios. We show that the situation is different for orbits close to the Galactic disc, whose vertical frequency $\Omega_z$ is strongly amplitude dependent. As a consequence, stars stripped from a satellite obtain a range of values for $\Omega_z$ and hence of frequency ratios, and spread into two dimensions, forming a ribbon-like structure with vertical extent comparable to that of the progenitor orbit. In integrals-of-motion space, tidal ribbons are clumps, which offers the best chance of detection and allows the determination of the Galactic potential vertically across the disc.
\end{abstract}


\section{Introduction}
Tidal debris streams form via the stripping of stars from galactic satellites by the tidal field of their host and the subsequent spreading of the clouds of stripped stars into one-dimensional structures within six-dimensional phase space. Dozens of such streams have been discovered in the Milky Way from two- or three-dimensional data alone \citep[e.g.][]{GrillmairCarlin2016}, but more are likely to be found by searches in more phase-space dimensions. The observable characteristics of these streams constrain the Galactic gravitational potential, including its time evolution, without the need to assume equilibrium of some tracer population. They have, for example, been proposed to constrain the distribution of dark sub-haloes orbiting the Milky-Way (\citealt*{YoonEtAl2011}; \citealt{ErkalEtAl2016}; \citealt*{BovyErkalSanders2017}), or used to infer the shape of the Galactic halo (e.g.\ \citealt*{KoposovRixHogg2010, BowdenBelokurovEvans2015}; \citealt{BovyEtAl2016}).

The orbits of debris stars differ only very little from that of their progenitor, and their drifting away from it originates from that difference. Thus, the debris delineates not a Galactic orbit, but the difference between nearby orbits, which accumulates over time. The orbits of the debris stars also differ between them and comprise a three-dimensional clump in orbit-space. Why then does the debris in general form only a one-dimensional structure?

The main reason is that the \emph{directions} in which nearby orbits diverge hardly depend on these orbital offsets (in contrast to their \emph{rates} of divergence). This is because for many Galactic potentials the \emph{ratios} of the orbital frequencies hardly vary. In other words, the frequency vectors have a preferred direction and their differences also point (almost) in this direction. A preferred direction of divergence from the progenitor implies a one-dimensional structure. Moreover, as this direction is close to that of an orbit, such a stream nearly delineates an orbit (though not that of the progenitor).

Mathematically, this is best described using action-angle variables (as first pointed out by \citealt{Tremaine1999, HelmiWhite1999}, see also Section~\ref{sec:action-angle}), which provide a conceptually clear and concise picture for the linear drift of the debris away from the progenitor. For a chaotic progenitor orbit, action-angle variables cannot be constructed and nearby trajectories diverge exponentially rather than linearly. This results in widening of the stream to a degree which depends on that of the chaos \citep{PriceWhelanEtAl2016}. In fact, if observed, such widening may be used to place constraints on the presence of chaos in the Galactic potential.

In this study, we show that debris from progenitors on Galactic disc orbits, i.e.\ with vertical excursion comparable to a few disc scale heights, forms not one- but two-dimensional structures: vertically extended ribbons around the Galaxy. Such debris ribbons can be very useful for constraining the vertical disc potential, i.e.\ for inferring the Oort limit and the total mass distribution vertically across the disc, because they trace the full vertical extent at each visited $(R,\,\phi)$ position within the disc, which may included the Solar neighbourhood. Here, we present numerical simulations demonstrating the formation of tidal ribbons and provide an explanation in terms of the action-angle formalism.

\section{Tidal debris from a disc progenitor}
\label{sec:simulations}
We simulated the tidal debris from a progenitor with a mass of $10^6\,\Msun$ orbiting an axisymmetric model of the Milky Way \citep{Allen1991} for $5\,$Gyr on the orbit of the globular cluster 47\,Tuc. Debris is represented as test particles initially sampled onto weakly bound orbits within the progenitor, thereby correctly modelling the stripping process (Hasanuddin \& Dehnen, in preparation).

\begin{figure}
	\includegraphics[height=49mm]{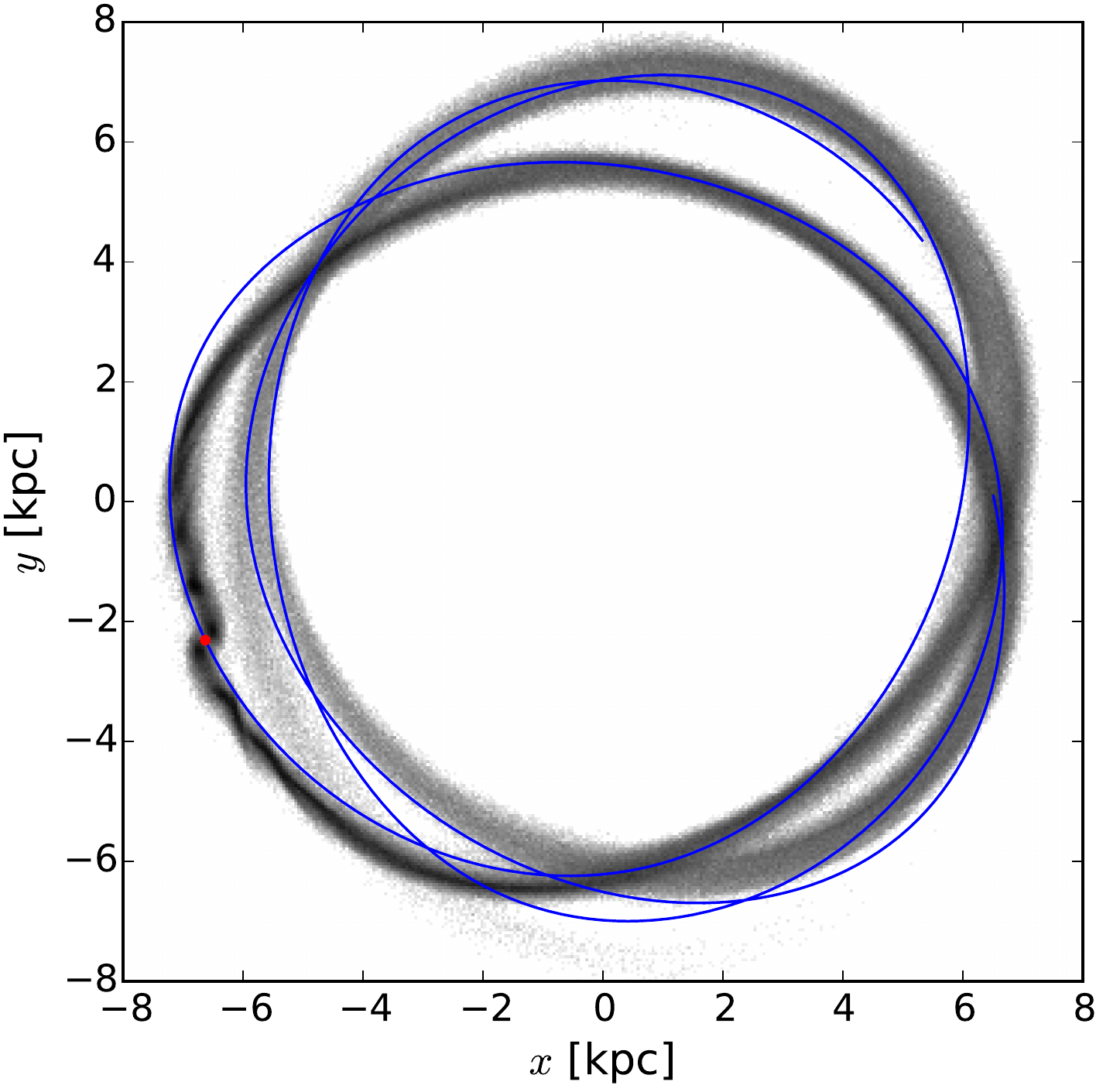}\hfill
	\includegraphics[height=49mm]{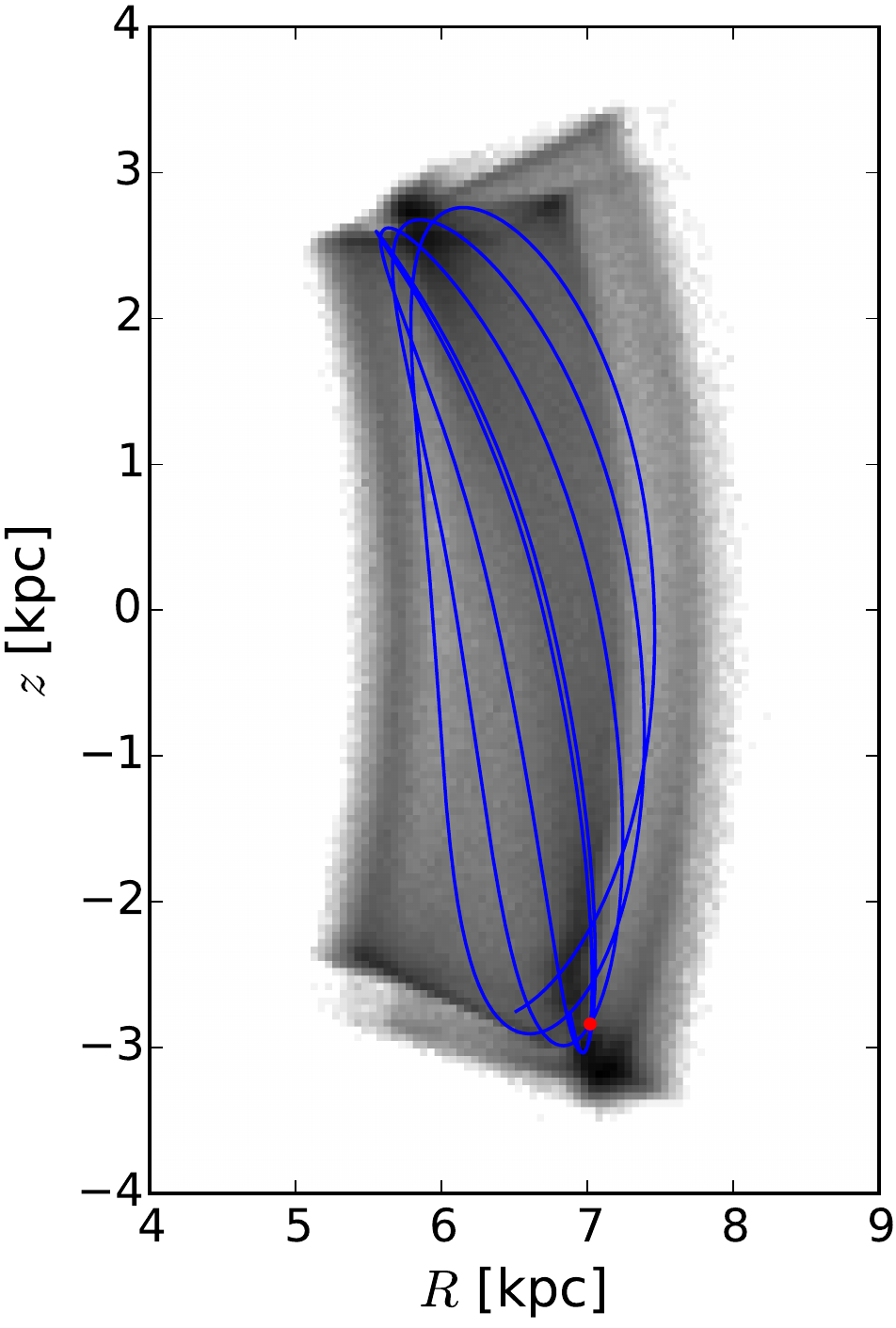}
	\vspace*{-1mm}
    \caption{Distribution of simulated debris formed by tidal stripping of stars from a globular cluster (red) orbiting the Milky way on a disc orbit (blue) in the $x$-$y$ (left) and $R$-$z$ (right) planes (note the different scales).}
    \label{fig:tails:XYZ}
\end{figure}

\begin{figure}
    \hfil\includegraphics[width=\columnwidth]{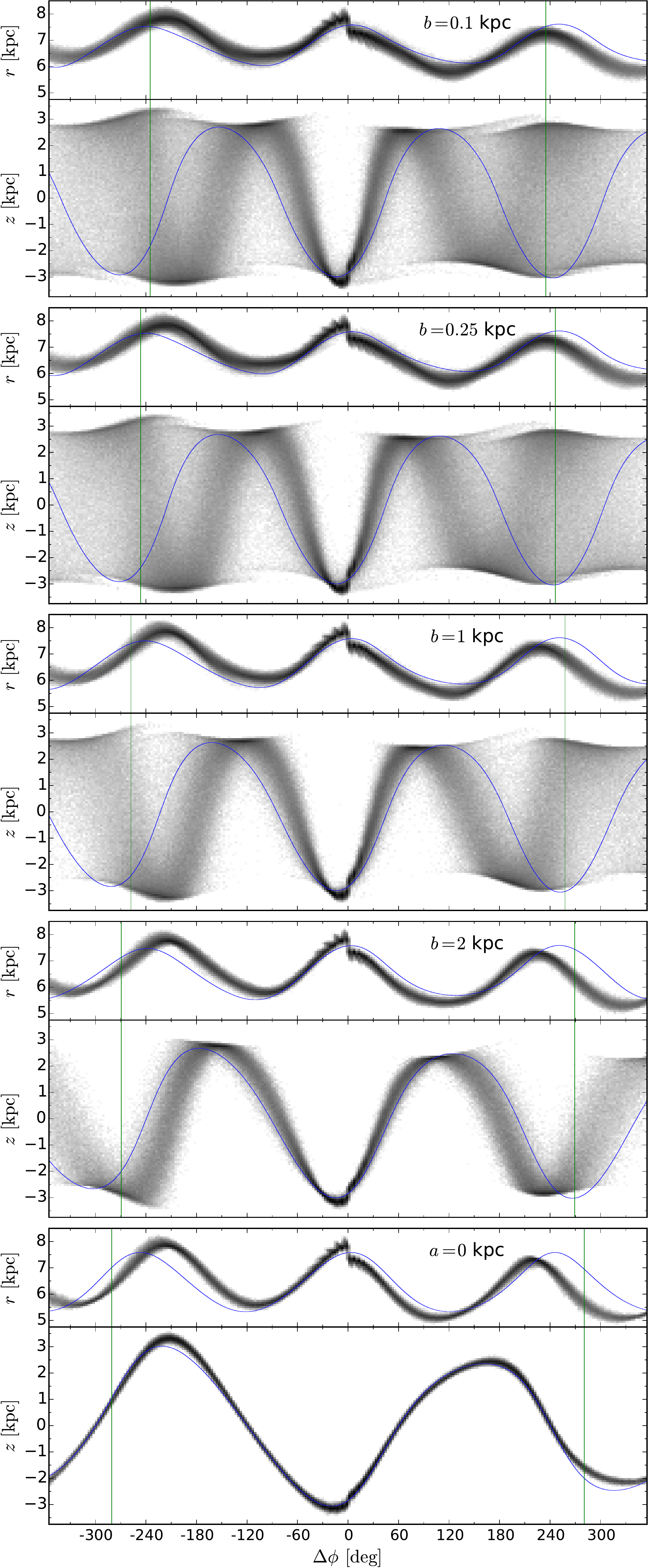}
    \vspace*{-4mm}
    \caption{Distribution of debris over azimuthal offset $\Delta\phi$ from the progenitor, spherical radius $r$, and vertical height $z$. The different panels correspond to varying degrees of flattening of the Galactic potential: flattest \emph{top} and spherical \emph{bottom}. The case $b=0.25$ is the same simulation as shown in Fig.~\ref{fig:tails:XYZ}. The models are only complete between the thin vertical lines: beyond the lines particles lost prior the start of the simulation would contribute.}
   \label{fig:tails:rphiz}
\end{figure}

Fig.~\ref{fig:tails:XYZ} plots projections of the spatial distribution of the resulting tidal debris. In the Galactic plane (left panel), the debris mimics an orbit, but vertically it resembles a band or ribbon. This is also evident in Fig~\ref{fig:tails:rphiz}, where the debris shows a narrow distribution over azimuthal offset $\Delta\phi$ and radius $r$, but spreads widely over $\Delta\phi$ and vertical position $z$. In fact, Fig~\ref{fig:tails:rphiz} shows the results from five simulations for different choices of the parameters $a$ and $b$ (as indicated) of the contribution \citep{Miyamoto1975}
\begin{equation}
    \sub{\Phi}{disc}(R,z) = -\frac{G\sub{M}{disc}}{\sqrt{R^2+\left(a+\sqrt{b^2+z^2}\right)^2}}
\end{equation}
of the Galactic disc to the gravitational potential, at fixed $a+b=5.32\,\kpc$. For $a=0$, this reduces to a spherical \cite{Plummer1911} model, and $b=0$ corresponds a razor-thin \cite{Kuzmin1956} disc. The potential in the disc mid-plane ($z=0$) only depends on the sum $a+b$, which acts as scale radius, such that Galactic models with different $a,\,b$ but the same $a+b$ have the same rotation curve.

For the spherical case ($a=0$, bottom panel of Fig~\ref{fig:tails:rphiz}) the debris forms a narrow stream, especially in the $\Delta\phi$-$z$ plot, where it closely resembles an orbit (this is an immediate consequence of sphericity when the vertical and azimuthal motions are just two aspects of inclined rotation). But for thinner discs (smaller $b$), the debris deviates from this picture and fans out to form a vertically extended band. This can be readily understood in terms of action-angle variables, as we now show. 

\section{Tidal debris in action-angle space}
\label{sec:action-angle}
For many situations in galaxy dynamics, the canonical action-angle phase-space coordinates provide the clearest insight, and tidal debris is no exception. Pioneered by \cite{Tremaine1999} and \cite{HelmiWhite1999}, many studies have used them to this purpose \citep[e.g.][]{EyreBinney2011, SandersBinney2013a, SandersBinney2013b, Sanders2014, Bovy2014} as we now summarize.

We denote the angle coordinates by $\vec{\theta}$ and their conjugate momenta, the actions, by $\vec{J}$. For regular disc orbits, the three actions are $J_r$, $J_\phi$, and $J_z$ \cite[for more details and a precise definition see][\S3.5]{BinneyTremaine2008}. The Hamiltonian is a function of the actions only: $H=H(\vec{J})$ and the canonical equations of motion
\begin{equation}
	\label{eqs:action-angle:motion}
	\diff{\vec{\theta}}t = \vec{\Omega}(\vec{J}) \equiv \pdiff{H}{\vec{J}}
	\qquad\text{and}\qquad
	\diff{\vec{J}}t = - \pdiff{H}{\vec{\theta}} = 0
\end{equation}
have the simple solutions $\vec{J}(t)=\vec{J}_{\!0}$ and 
\begin{equation}
	\label{eq:theta:t}
	\vec{\theta}(t) = \vec{\theta}_0 + t\,\vec{\Omega},
\end{equation}
where a subscript `0' denotes initial values \citep[e.g.][]{Goldstein1980}.

The initial angles and actions of debris stars differ only very little from those of their progenitor. The corresponding offsets, $\Delta\vec{\theta}_0$ and $\Delta\vec{J}$, then fully describe the time evolution of the debris relative to the progenitor. From the equations~\eqref{eqs:action-angle:motion} of motion, we find that $\Delta\vec{J}$ remains at its initial value, while at time $\Delta t$ after the stripping
\begin{equation}
	\label{eq:Delta:theta:t}
	\Delta\vec{\theta}(t) = \Delta\vec{\theta}_0 + \Delta t\,\Delta\vec{\Omega}
	\approx  \Delta t\;\Delta\vec{\Omega}.
\end{equation}
A comparison with equation~\eqref{eq:theta:t} shows that the debris behaves like an orbit, but with the angles and frequencies, $\vec{\theta}$ and $\vec{\Omega}$, replaced with the respective differences, $\Delta\vec{\theta}$ and $\Delta\vec{\Omega}$, between debris star and progenitor. Since
\begin{equation}
	\label{eq:DeltaJ:M:third}
	\Delta\vec{J}\cdot\vec{\Omega} \approx \Delta E \approx \Delta\vec{x}\cdot\vec{\nabla}\Phi \propto \sub{r}{tid}|\vec{\nabla}\Phi| \propto \sub{M}{sat}^{1/3},
\end{equation}
\citep[e.g.][]{Johnston1998,DehnenEtal2004,SandersBinney2013a} $\Delta\vec{J}\ll\vec{J}$ for debris from small satellites, and we can accurately approximate the frequency offset from its Taylor expansion
\begin{equation}
	\label{eq:Delta:Omega}
	\Delta\vec{\Omega} \approx \mathbfss{H} \cdot \Delta\vec{J}
\end{equation}
with the Hessian of the Hamiltonian \citep{Tremaine1999}
\begin{equation}
	\mathsf{H}_{ik} \equiv \ptdiff{H}{J_i\,}{J_k}.
\end{equation}
This formalism separates the orbital dynamics, encapsulated in the matrix $\mathbfss{H}$ (and the canonical map from $\vec{\theta},\,\vec{J}$ to Cartesian phase-space co-ordinates), from the stripping process, providing the distribution over $\Delta\vec{\theta}_0$, $\Delta\vec{J}$ and $\Delta t$. Tidal streams, shells, and ribbons can all be naturally explained in this way, as we now demonstrate.

\subsection{When does debris form a thin stream?}
\label{sec:stream}
Consider the situation of a Hamiltonian that depends on the actions only through a linear combination:
\begin{equation}
	\label{eq:H:planar}
	H = f(I),\qquad I=\vec{\omega}\cdot\vec{J}
\end{equation}
with a constant vector $\vec{\omega}$ and some function $f$. In this case, energy surfaces in action space are parallel planes, $\vec{\Omega}=f'(I)\,\vec{\omega}$, and
\begin{equation}
	\label{eq:Hessian:Om:Om}
	\mathsf{H}_{ik} \propto \Omega_i\,\Omega_k
\end{equation}
is proportional to the outer self-product of $\vec{\Omega}$. Such a $\mathbfss{H}$ has only one non-zero eigenvalue $\lambda_1$ with eigenvector $\hat{\vec{e}}_1\propto\vec{\Omega}$ \citep{Tremaine1999}. Therefore, \emph{any} action offset $\Delta\vec{J}$ is projected onto $\Delta\vec{\Omega} \propto \vec{\Omega}$, and tidal debris forms one-dimensional streams aligned to an orbit. 

For a spherical galaxy, $H$ depends on $J_\phi$ and $J_z$ only through the total angular momentum $L=|J_\phi|+J_z$ and one need only consider the actions $J_r$ and $L$. For scale-free spherical galaxies, i.e.\ with power-law circular-velocity $\sub{v}{c}(r)\propto r^\beta$, 
\begin{equation}
	\label{eq:J:first}
	J_r(E,L) \approx \varphi^{-1}\left[\sub{L}{c}(E) - L\right]
\end{equation}
\citep{Dehnen1999:Epicycle} with $\varphi=\kappa/\omega=\sqrt{2(\beta+1)}$ the ratio of the epicycle frequencies and $\sub{L}{c}(E)$ the angular momentum of the circular orbit with energy $E$. The implied $H(\vec{J})$ is of the form~\eqref{eq:H:planar} with $\vec{\omega}=(\varphi,1,1)$, such that the frequencies are approximated as those of the circular orbit with the same energy, and streams as thin and orbit-aligned \citep*{WilliamsEtAl2014}.

For the harmonic ($\beta=1$) and point-mass potentials ($\beta=-\tfrac12$) this approximation becomes exact: tidal-disruption events near super-massive black holes form elliptic streams (barring general relativistic effects), as does debris from comets or asteroids (barring the effect of the Solar wind). However, for $-\tfrac12<\beta<1$ equation~\eqref{eq:J:first} slightly over-estimates $J_r$ at small eccentricities
\begin{align}
	e \equiv \sqrt{1-L^2/\sub{L}{c}^2(E)},
\end{align}
with the largest deviation for $\beta=0$ \citep[flat rotation curve; see][Fig.~9]{Dehnen1999:Epicycle}. This deviation from the form~\eqref{eq:H:planar} implies that the Hessian $\mathbfss{H}$ has a second non-vanishing eigenvalue $\lambda_2$, and that the eigenvector $\hat{\vec{e}}_1$ no longer points exactly in the direction of $\vec{\Omega}$, but deviates by some angle $\vartheta$.

Hence, an isotropic distribution of $\Delta\vec{J}$ results in an anisotropic distribution of $\Delta\vec{\Omega}$ with axis ratio $|\lambda_2/\lambda_1|$.
In reality, $\Delta\vec{J}$ is bi-modally distributed as stars stripped from the inner and outer Lagrange points have $\Delta E<0$ and $>0$, respectively, and directions resulting in $\Delta E=\vec{\Omega}\cdot\Delta\vec{J}=0$ are avoided (see also \citeauthor{EyreBinney2011}, \S6). This increases the anisotropy of the $\Delta\vec{\Omega}$ distribution, but not by much (according to our numerical experiments). This anisotropy directly translates into a relative divergence rate for the debris, implying (assuming the order $|\lambda_1|\ge|\lambda_2|\ge|\lambda_3|$)
\begin{equation}
	\label{eq:thin:stream:condition}
	|\lambda_1|\gg|\lambda_2|
\end{equation}
as condition for the formation of thin streams\footnote{But not $\lambda_1\gg\lambda_2$ \citep{EyreBinney2011, SandersBinney2013a} as typically $\lambda_1<0$ because the frequencies decrease with increasing actions.}. Such a stream spreads out in the direction $\hat{\vec{e}}_1$, and the angle $\vartheta$ between $\hat{\vec{e}}_1$ and $\vec{\Omega}$ is also the angle in $\vec{\theta}$ space between the stream and an orbit (\citeauthor{EyreBinney2011}, \citealt{SandersBinney2013a}).

We now estimate $|\lambda_2/\lambda_1|$ and $\vartheta$ for orbits in scale-free spherical galaxies by going beyond the plane-parallel approximation. To this end, we note that the radial frequency is very well approximated by the second-order expression \citep{Dehnen1999:Epicycle}
\begin{equation}
	\Omega_r(E,L) = \frac{\kappa(E)}{1-\zeta e^2}
	\quad\text{with}\quad\zeta \equiv \tfrac12(2-\varphi)(\varphi-1)\varphi^{-2}
\end{equation}
at all eccentricities. This corresponds to
\begin{align}
	\label{eqs:Jr:second}
	J_r &= \varphi^{-1} [\sub{L}{c}-L]\left(1 - \zeta\sub{L}{c}^{-1}[\sub{L}{c}-L]\right),
\end{align}
which extends the linear approximation~\eqref{eq:J:first} by a term quadratic in $\sub{L}{c}-L$. The implied Hamiltonian is a function of the non-linear combination 
\begin{align}
	\label{eq:H:curved}
	I &= \frac{1}{2(1-\zeta)} \left[\varphi J_r+(1-2\zeta)L
		+ \sqrt{\varphi^2J_r^2 + 2\varphi(1-2\zeta)J_r L + L^2}\right]
\end{align}
of the actions $J_r$ and $L$, and has derivatives
\begin{subequations}
\label{eqs:H:power:law:approx}
\begin{align}
	\pdiff{H}{J_r} &= \Omega_r
		= \frac{\kappa}{1-\zeta e^2},
	\\[0.5ex]
	\pdiff{H}{L} &= \Omega_\phi
		= \frac{\omega}{1-\zeta e^2}
			\left[1-2\zeta\left(1-\sqrt{1-e^2}\right)\right],
	\\[0.5ex]
	\label{eq:HJJ:power:law:approx}
	\ptdiffx{H}{J_r}
		&= -\left[\frac{1-\beta}{1+\beta} -
			\frac{2\zeta(1-e^2)}{1-\zeta e^2}\right]
			\frac{\Omega_r^2}{\omega L_c},
	\\[0.5ex]	
	\ptdiff{H}{J_r}L
		&= 	-\left[\frac{1-\beta}{1+\beta} - 
			\frac{2\zeta(1-e^2)}{1-\zeta e^2}\right]
			\frac{\Omega_r\Omega_\phi}{\omega L_c}
			-\frac{2\zeta\sqrt{1-e^2}\,\Omega_r}{(1-\zeta e^2)\,\sub{L}{c}},
	\\[0.5ex]
	\ptdiffx{H}{L} 
		&=  -\left[\frac{1-\beta}{1+\beta} -
			\frac{2\zeta(1-e^2)}{1-\zeta e^2}\right]
			\frac{\Omega_\phi^2}{\omega L_c}
			-\frac{2\zeta(2\sqrt{1-e^2}\,\Omega_\phi-\omega)}{(1-\zeta e^2)\,\sub{L}{c}}.
\end{align}
\end{subequations}
Thus, the Hessian consists of a first part proportional to $\Omega_i \Omega_j$, as for plane-parallel energy surfaces, and a second part that deviates from this pattern. We find that the resulting approximations for the ratio $|\lambda_2/\lambda_1|$ and the angle $\vartheta$ are weak functions of $\beta$ and $e$, staying small ($\lesssim0.03$ and $\lesssim3^\circ$, respectively) for all $\beta$ and $e$.

The largest values for $|\lambda_2/\lambda_1|$ and $\vartheta$ occur for circular orbits, for which the exact $\mathbfss{H}$ can be obtained via extended epicycle theory for any spherical potential (see Appendix~\ref{app:A}). We find that for $e=0$ equations~\eqref{eqs:H:power:law:approx} are correct, except \eqref{eq:HJJ:power:law:approx}, which makes a maximal error of 4\% (at $\beta=0.15$). The implied errors of $|\lambda_2/\lambda_1|$ and $\vartheta$ are small, and we conclude that for spherical galaxies with near-flat rotation curves $|\lambda_2/\lambda_1|\lesssim0.03$ and $\vartheta\lesssim3^\circ$.

These results quantify the statement by \cite{WilliamsEtAl2014} that streams in scale-free spherical systems are thin and nearly orbit-aligned. This also agrees with the findings of \cite{BowdenBelokurovEvans2015} that modelling the Galactic stream GD-1 using an orbit or a proper stream model makes hardly any difference in a logarithmic potential ($\beta=0$), even if that is oblate. 

Moreover, even without knowledge of $H(\vec{J})$, \cite*{ErkalSandersBelokurov2016} have shown that debris from progenitors on loop orbits in oblate or triaxial galaxies diverges faster with increasing asphericity or orbital inclination. However, the divergence rates are still modest and result merely in a slow widening of the stream, implying that  condition~\eqref{eq:thin:stream:condition} is still satisfied, albeit not to the same degree as for loop orbits in spherical galaxies.

\subsection{When does debris form a radial shell?}
\label{sec:shell}
Another well-known form of tidal debris are radial shells, observed in $\sim20\%$ of elliptical galaxies \citep{TalEtAl2009} and found in simulations of the tidal disruption of satellites on plunging orbits \citep[e.g.][]{HernquistQuinn1988, HernquistQuinn1989}. These shells differ from streams in that the debris spreads over a wide range of directions, such that the accumulation near apo-centre causes a shell-like appearance.

\begin{figure}
    \includegraphics[width=\columnwidth]{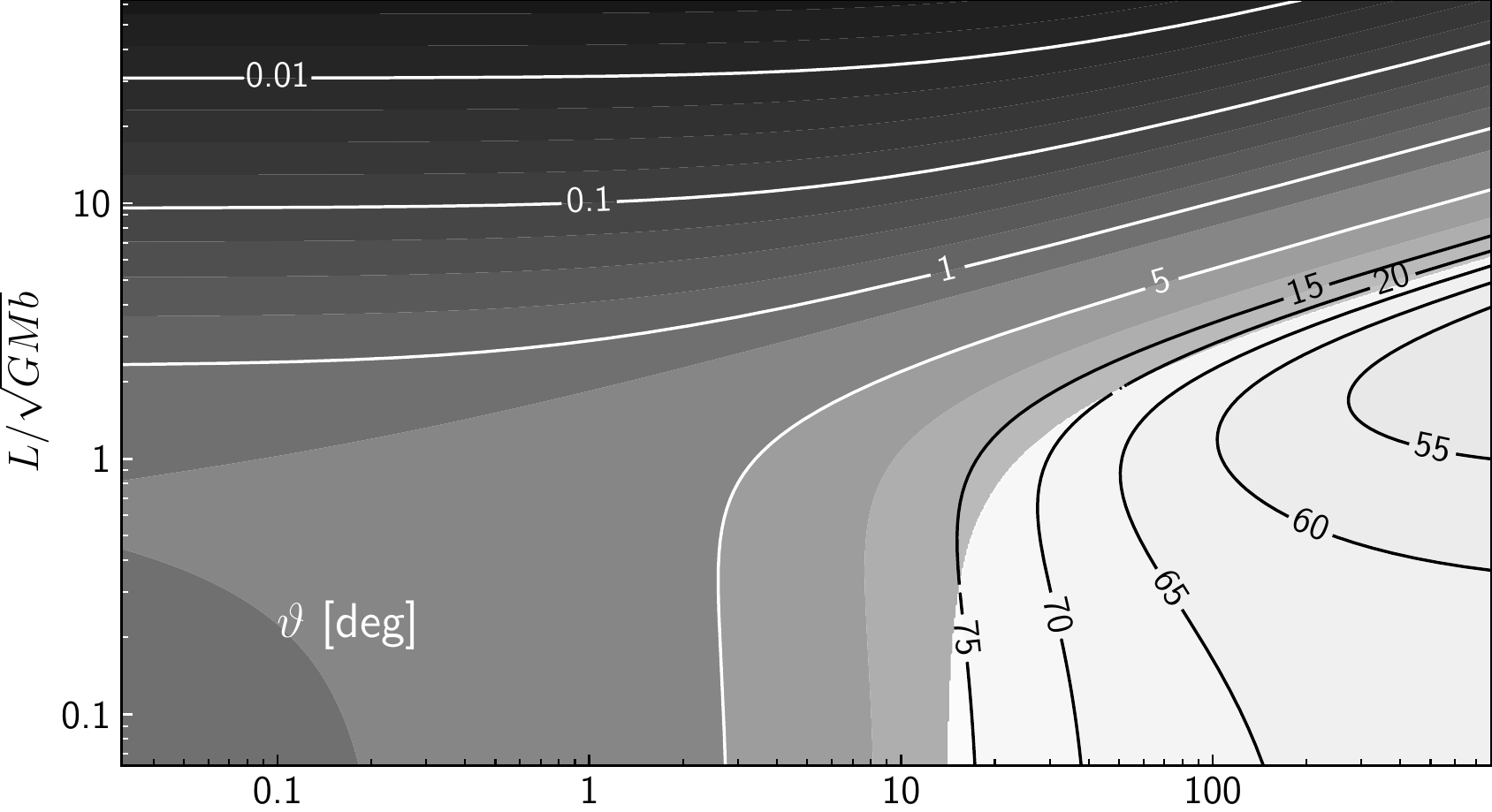}

    \includegraphics[width=\columnwidth]{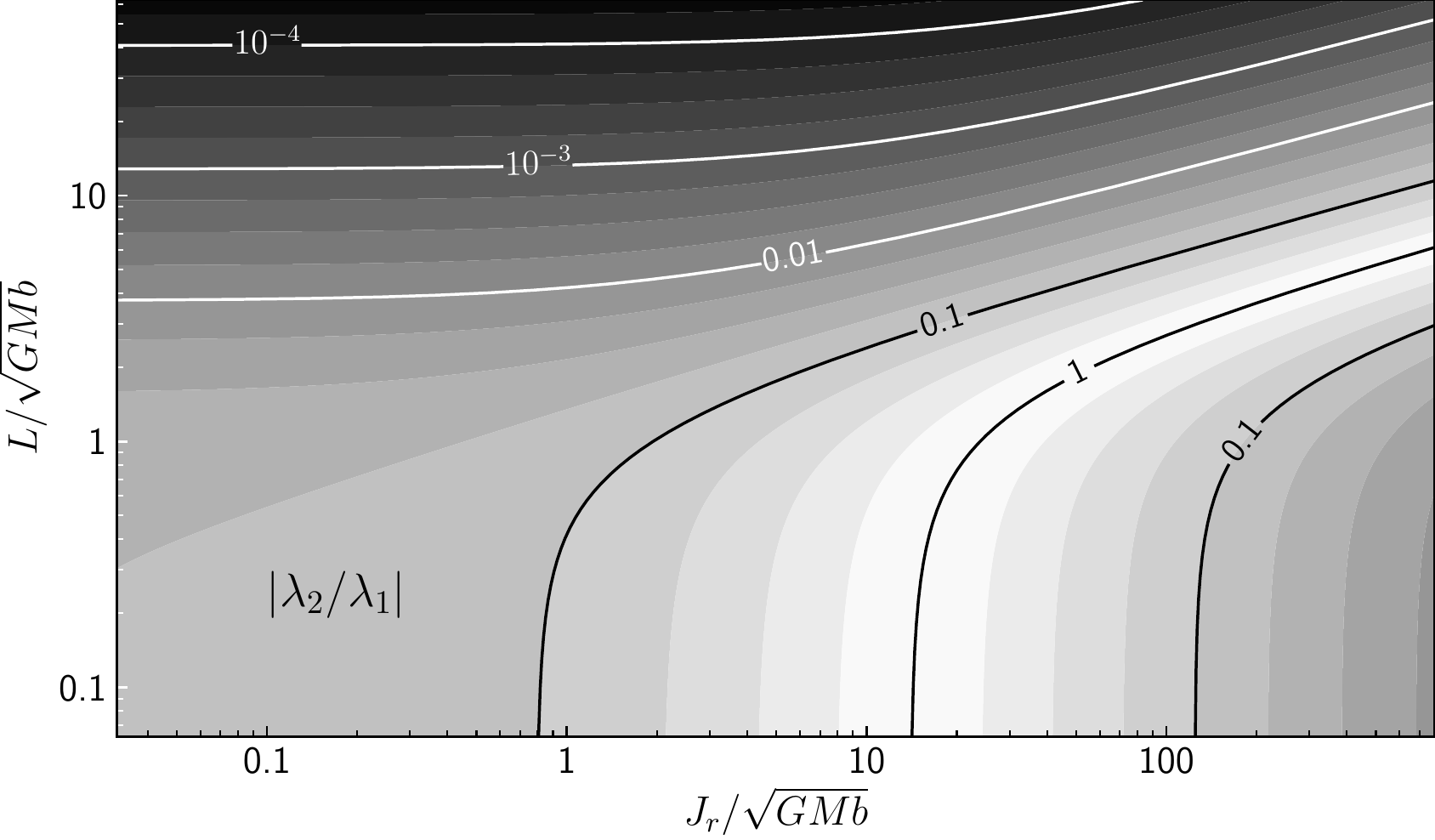}
    \vspace*{-4mm}
    \caption{Contours of $\vartheta$ (the angle between $\hat{\vec{e}}_1$ and $\vec{\Omega}$) and $|\lambda_2/\lambda_1|$ in action space for the isochrone potential\setcounter{footnote}{1}\protect\footnotemark. To the left of the $|\lambda_2/\lambda_1|=1$ contour $\lambda_1<0<\lambda_2$, while to its right $\lambda_2<0<\lambda_1$ and $\vartheta$ changes by $90^\circ$ when crossing between these regions. Debris from orbits with $|\lambda_2|\sim|\lambda_1|$ forms two-dimensional structures.}
    \label{fig:ratio:iso}
\end{figure}

Such shells are also described by the action-angle formalism. Consider, for example, the galaxy potential known as \citeauthor{Henon1959}'s (\citeyear{Henon1959}) isochrone sphere (with $b$ a scale radius)
\begin{equation}
	\Phi(r) = -\frac{GM}{b+\sqrt{r^2+b^2}}.
\end{equation}
While this model is not very realistic, since it implies a constant-density core and a $\rho\propto r^{-4}$ envelope, it has the benefit that 
\begin{equation}
	H(J_r,L) = -\tfrac12 G^2M^2 \left[J_r+\tfrac12\left(L+\sqrt{4GMb+L^2}\right)\right]^{-2}
\end{equation}
\citep{Saha1991} is known analytically. We have (e.g.\ \citeauthor{Saha1991})
\begin{subequations}
\begin{equation}
	\Omega_r = \sqrt{\frac{GM}{a^3}},\quad
	\Omega_\phi = \frac{\Omega_r}{2} \left[1+\frac{L}{\sqrt{L^2+4GMb}}\right]
\end{equation}
with $a\equiv-GM/2H$ and the Hessian follows as \citep[see also][]{Eyre2010PhD}\!\!
\begin{equation}
	\mathbfss{H} = -\frac{3}{a^2}
	\left(\begin{array}{ll}
		1 & g \\
		g\; & g^2
	\end{array}\right)
		+ \sqrt{\frac{GM}{a^3}}
	\left(\begin{array}{ll}
		0 & 0 \\
		0\; & g'
	\end{array}\right),
\end{equation}
\end{subequations}
where $g=g(L)\equiv\Omega_\phi/\Omega_r$. Deviations of $\mathbfss{H}$ from the form~\eqref{eq:Hessian:Om:Om} are due to the second term, which has the opposite sign (as $g'>0$) and dominates for weakly bound near-radial orbits. Such orbits probe both the Keplerian envelope and the harmonic core, and have $\partial \Omega_\phi/\partial L>0$, reversing the usual decrease of frequency with increasing actions. If $\partial \Omega_\phi/\partial L + \partial\Omega_r/\partial J_r=0$, then $\lambda_1=-\lambda_2$, see Fig.~\ref{fig:ratio:iso}\footnote{\citeauthor{EyreBinney2011} (\citeyear{EyreBinney2011}, Fig.~3) plot contours of $\lambda_1/\lambda_2$ and $\vartheta$ for the isochrone potential, but only for $J_r<0.5\sqrt{GMb}$ and $L<2\sqrt{GMb}$, the bottom left corners of our plots. They incorrectly report $\vartheta<2.\!\!\!^\circ2$ and $\lambda_1/\lambda_2>10$ everywhere (in fact $\lambda_1/\lambda_2<0$) and missed that $|\lambda_1|\sim|\lambda_2|$ for some orbits.}, and debris spreads equally in two directions forming two-dimensional structures, the caustics of which appear as radial shells.

This situation is quite different from that for scale-free models, and may not explain radial shells in cusped galaxies. However, for plunging orbits in a cusped galaxy, we still expect radial shells if $\Delta{L}\gtrsim L$. In this case, the approximation~\eqref{eq:Delta:Omega} is invalid \citep{EyreBinney2011} and with it all reasoning based on the Hessian of $H(\vec{J})$, in particular the result from Section~\ref{sec:stream} that debris in scale-free potentials always forms thin streams. For purely radial orbits ($L=0$), debris stars will pass the centre of the Galaxy on all sides and get deflected in all directions (with deflection angle $\neq0$ for radial orbits in scale-free cusps), creating a wide distribution but with very similar apo-centres and hence forming radial shells.

For box orbits in triaxial systems, which regularly come close to the centre, a similar mechanism may apply, but understanding debris from box orbits requires the associated $\mathbfss{H}$ to be calculated, for example, for triaxial St\"ackel potentials.

\subsection{When does debris form a ribbon?}
\label{sec:ribbon}
\begin{figure}
    \hfil\includegraphics[width=\columnwidth]{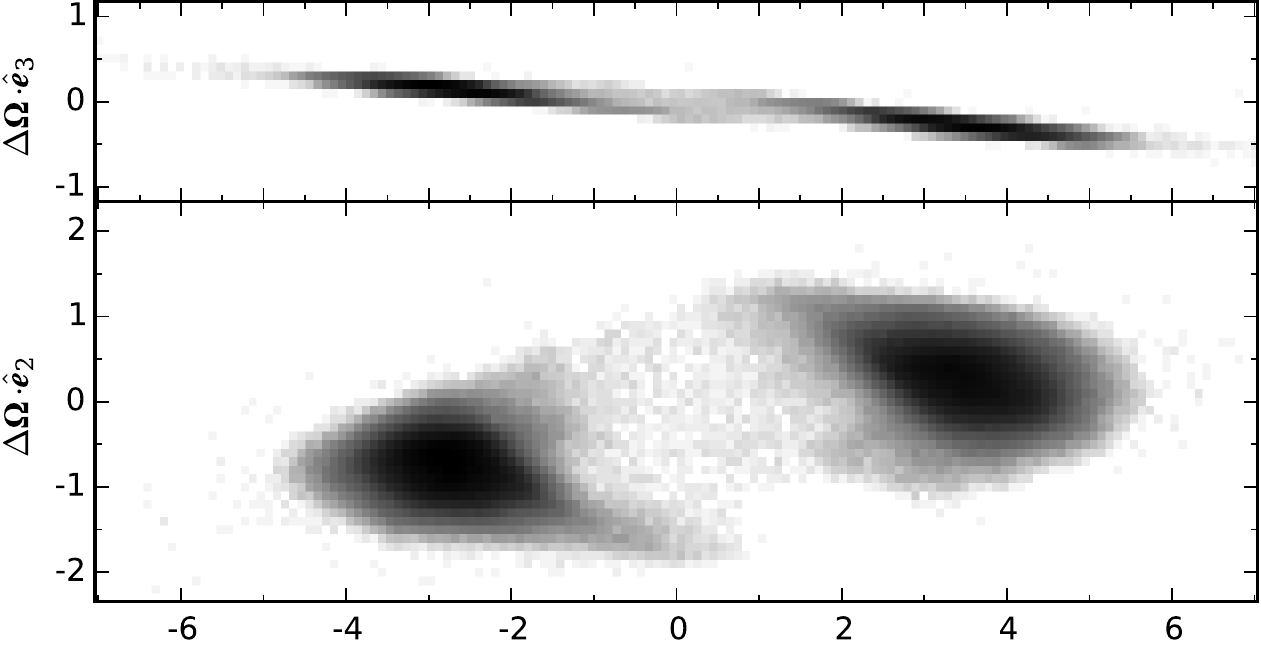}

    \hfil\includegraphics[width=\columnwidth]{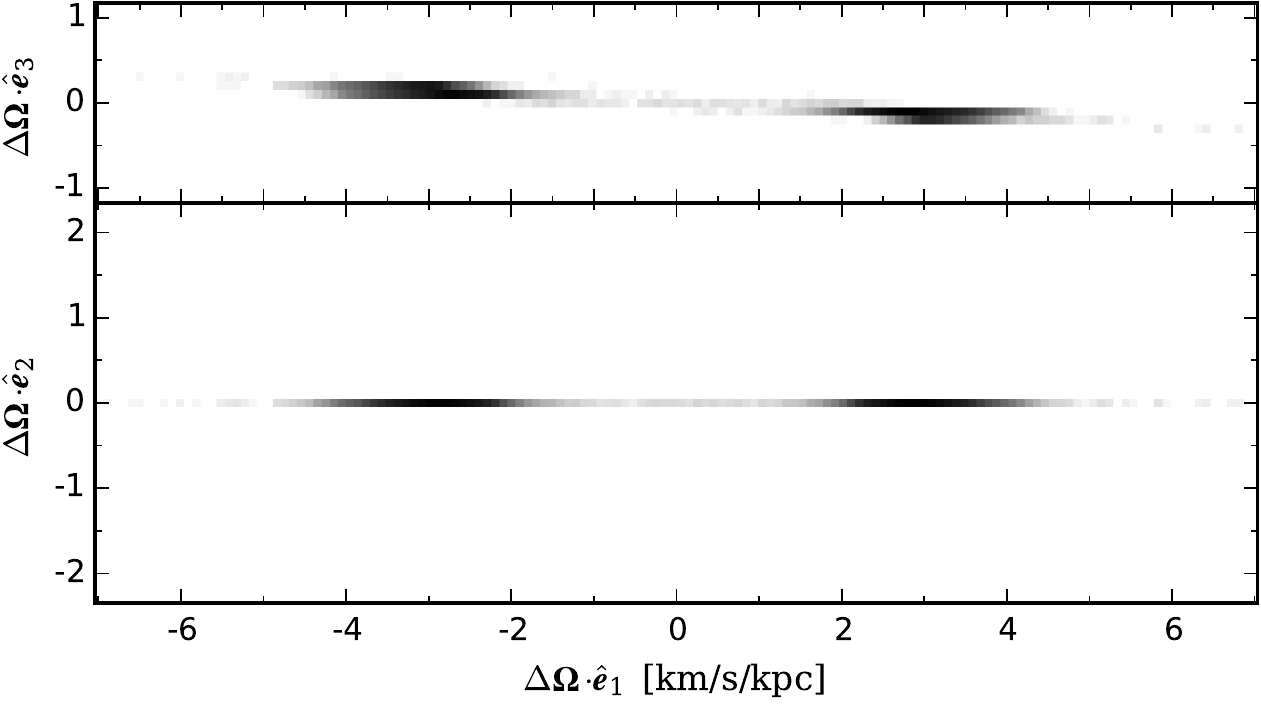}
    \vspace*{-4mm}
    \caption{\emph{Top}: Distribution of debris stars from a disc orbit (the simulation presented in Fig.~\ref{fig:tails:XYZ}) over $\Delta{\vec{\Omega}}$ projected onto the eigenvectors of the Hessian as numerically calculated (see text). \emph{Bottom}: similar, but for a simulation in a spherical galaxy (see bottom panel of Fig.~\ref{fig:tails:rphiz} for the spatial distribution).}
    \label{fig:tails:Delta:Om}
\end{figure}
While radial shells are caused by the violation of the condition~\eqref{eq:thin:stream:condition} on plunging orbits, ribbons occur when this condition is violated for loop orbits. Our simulations suggest that this is the case for disc orbits, but are not open to analytical insight. For that we now construct a toy model. On near-circular disc orbits, the vertical gravitational force is dominated by the attraction of the local Galactic disc. If we approximate the disc as razor thin, we have
\begin{equation}
	\label{eq:Phi:disc}
	\Phi(R,z) \approx \Phi(R,0) + 2\pi G\Sigma(R) |z|
\end{equation}
with $\Sigma(R)$ the surface mass density of the disc. For the one-dimensional vertical motion in this potential
\begin{align}
	\label{eq:Ez:Oz:Jz}
	E_z = \tfrac12 (3\pi^2G\Sigma\,J_z)^{2/3}\qquad\text{and}\qquad
	\Omega_z = \diff{E_z}{J_z} \propto J_z^{-1/3}.
\end{align}
The most important property of these relations is the non-linear dependence on the vertical action $J_z$. Even if the assumptions made here are not fully correct, the basic fact that $\Omega_z$ is a strong function of $J_z$ for orbits with vertical extent in the range $\sim\,$0.4-4\,\kpc\ remains valid. If we ignore the slight curvature of the energy surfaces in the $J_r$-$J_\phi$ plane (which we discussed in Section~\ref{sec:stream}), we expect, for example from a Hamiltonian of the form $H=f(\varphi J_r+|J_\phi|+j^{1/3}J_z^{2/3})$ with appropriate constant $j$, that the Hessian is approximately
\begin{align}
	\mathsf{H}_{ik} \propto \Omega_i\,\Omega_k + \alpha_i\alpha_k
\end{align}
with $\vec{\alpha}\sim\Omega_z\hat{\vec{z}}$. For $\Omega_r:\Omega_\phi:\Omega_z=1.5:1:1.4$, this Hessian obtains $|\lambda_2/ \lambda_1|=0.25$ and $\hat{\vec{e}}_2\cdot\vec{\Omega}/|\vec{\Omega}|=0.46$, i.e.\ a substantial second eigenvalue with eigenvector far from orthogonal to $\vec{\Omega}$, and hence likely to be projecting onto most debris' $\Delta\vec{J}$. This agrees with the findings of \cite{SandersBinney2013a}, who numerically found $|\lambda_2/ \lambda_1|\gtrsim\tfrac18$ (read off their Fig.~2) for low-$J_z$ orbits in a realistic Galaxy model.

For the simulation presented in Section \ref{sec:simulations}, we numerically estimated the actions and frequencies for the progenitor and debris stars using \citeauthor{Bovy2015}'s (\citeyear{Bovy2015}) \textsc{galpy} code employing the St\"{a}ckel approximation \citep{Binney2012}, giving 4\% and 0.4\% rms deviations for $J_r$ and $J_z$ from their respective orbital means. From these, we obtain $\mathbfss{H}$ via a linear fit. The results of this exercise,
\begin{subequations}
\begin{align}
	\label{eq:unit:om:vec}
	J_{r,\phi,z} &\approx 13,\; 1243,\; 180\;\kpc\,\kms,\\[-0.1ex]
	\Omega_{r,\phi,z} &\approx 44,\;30,\;41\;\kms\,\kpc^{-1},\\[-0.1ex]
	\lambda_{1,2,3} &\approx -0.106,\;-0.033,\;+0.0011\;\mathrm{kpc}^{-2},\\[-0.1ex]
	\hat{\vec{e}}_{1,2,3} &\approx
		\begin{pmatrix} 0.43 \\ 0.37 \\  0.82\end{pmatrix},\quad
		\begin{pmatrix} \phm0.73 \\ \phm0.38 \\ -0.56    \end{pmatrix},\quad
		\begin{pmatrix} \phm0.52\\ -0.84\\ \phm0.11\end{pmatrix}, \\[-0.1ex]
	|\hat{\vec{e}}_{1,2,3}\cdot\hat{\vec{\Omega}}| &\approx 0.95,\;0.3,\;0.03,
\end{align}
\end{subequations}
are not far from our toy model.

In the top panels of Fig.~\ref{fig:tails:Delta:Om}, we project the numerically obtained $\Delta\vec{\Omega}$ for the simulated debris onto the eigenvectors $\hat{\vec{e}}_i$ of the Hessian. Evidently, the debris samples two planar regions, one for the leading and another for the trailing stream, in three-dimensional $\Delta\vec{\Omega}$ space, largely perpendicular to $\hat{\vec{e}}_3$. As a result, the debris does not form a thin stream but fans out vertically into a broader structure \citep{SandersBinney2013a}, clearly visible in Fig.~\ref{fig:tails:Delta:Om} near the progenitor. However, eventually the debris densely occupies a two-dimensional region in $\Delta\vec{\theta}$, though for typical debris stars it takes $\sim3$ times longer to wrap around in the $\hat{\vec{e}}_2$ than the $\hat{\vec{e}}_1$ direction.

In contrast, the bottom panels of Fig.~\ref{fig:tails:Delta:Om} show the situation for a spherical galaxy, when the distribution of $\Delta\vec{\Omega}$ is one-dimensional and the debris forms a thin stream (Fig.~\ref{fig:tails:rphiz}, bottom panel).

\section{Discussion}
\label{sec:discuss}
\begin{figure}
    \hfil\includegraphics[width=\columnwidth]{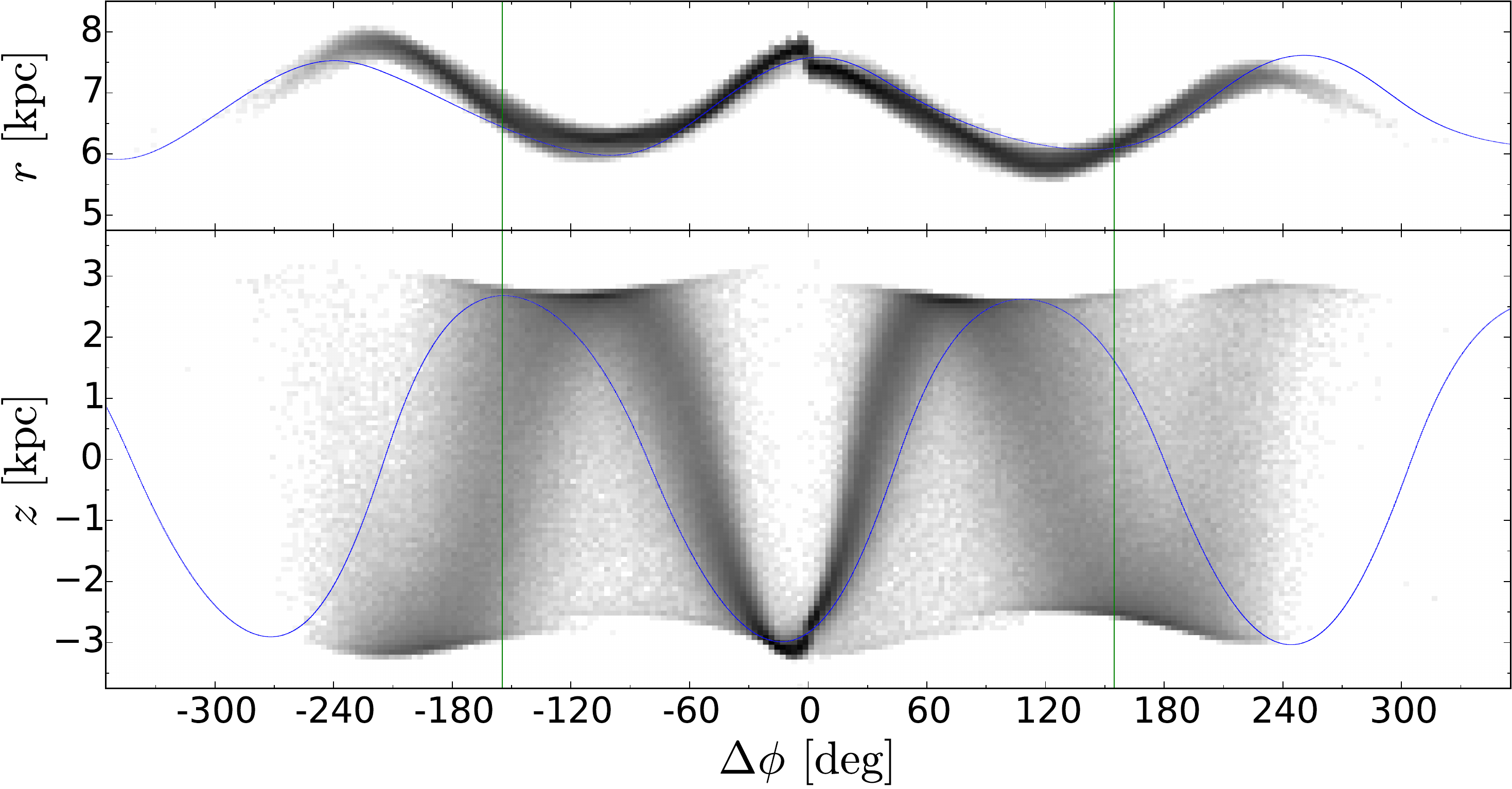}
    \vspace*{-3mm}
    \caption{As the panels for the $b=0.25$\,kpc case in Fig.~\ref{fig:tails:rphiz}, but for a progenitor of only $2\times10^5\Msun$, five times less massive.}
    \label{fig:tails_02Mc}
\end{figure}
We have demonstrated numerically and explained analytically the formation process of tidal ribbons, which are produced by tidal debris from progenitors on Galactic orbits with highly anharmonic vertical motion, which includes all disc orbits with vertical extent in the range $\sim$0.4-4$\,\kpc$. This also implies that extended tidal streams cannot exist near such orbits.

This phenomenon is entirely caused by the properties of the Galactic gravitational field (and implied Hamiltonian) and independent of the properties of the disrupted satellite. A less massive satellite on the same orbit, for example, obtains tidal ribbons with exactly the same spatial structure as demonstrated in Fig.~\ref{fig:tails_02Mc}, albeit with a drift rate reduced $\propto\sub{M}{sat}^{1/3}$, because of relation~\eqref{eq:DeltaJ:M:third}. However, there may be insufficient time for the formation of fully developed ribbons from small satellites, because even after another 5\,Gyr the debris will have wrapped just once around the Galaxy and it takes $\sim 3$ times longer for most debris stars to wrap around vertically.

These ribbons are two-dimensional entities in six-dimensional phase space and hence spread the debris over a larger region than conventional tidal streams, and could be hard to detect by stream-finding methods. However, a ribbon crossing the wider Solar vicinity should be detectable by exploiting that the debris stars have very similar actions or, equivalently, angular momentum as well as vertical and horizontal orbital energies. This approach is essentially the same as the `integrals of motion space' search for halo substructure \citep{HelmiEtAl1999, HelmiEtAl2017, MyeongEtAl2017}.

\begin{figure}
    \hfil\includegraphics[width=\columnwidth]{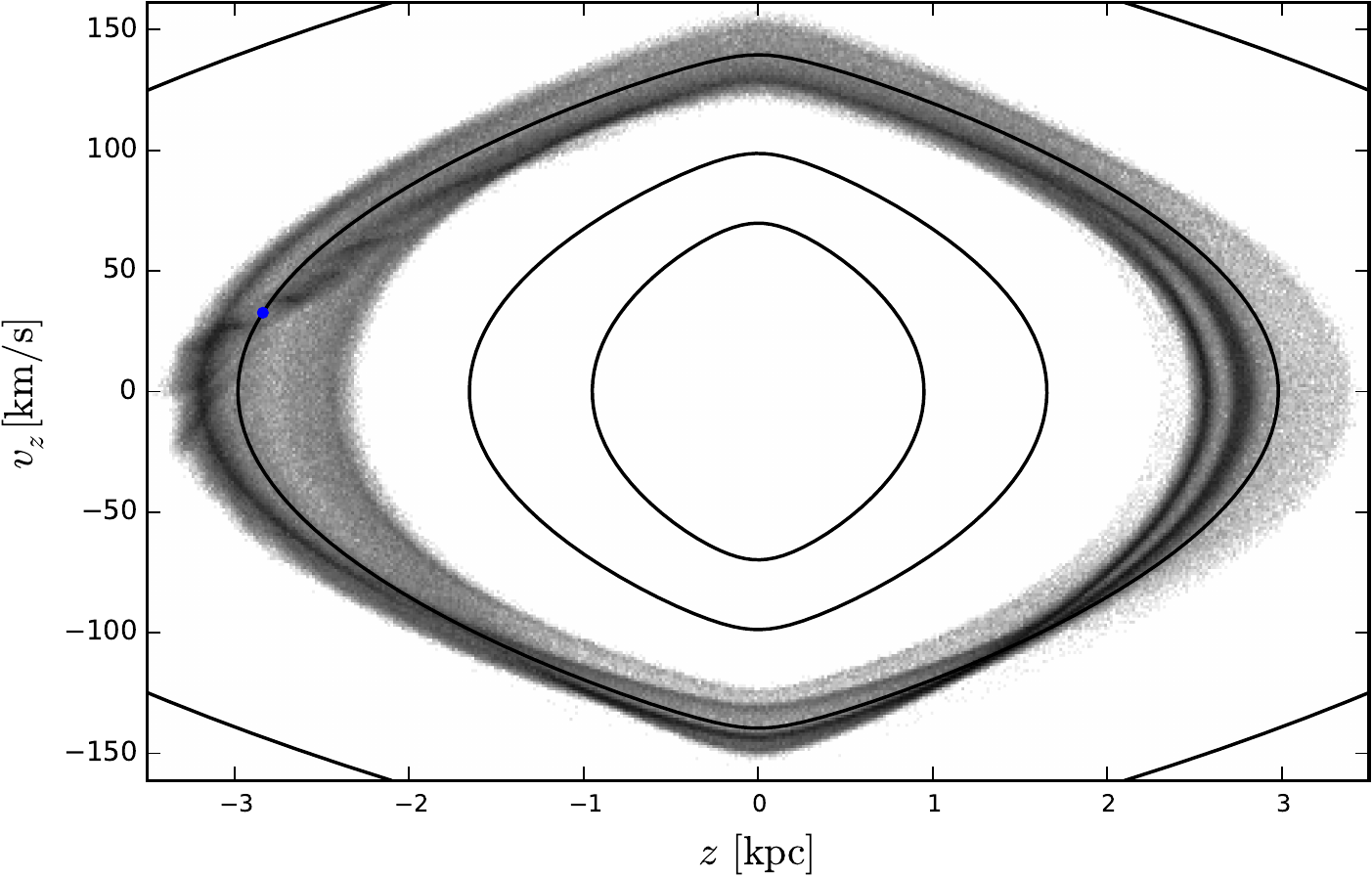}
    \vspace*{-4mm}
    \caption{Distribution of the simulated debris shown in Fig.~\ref{fig:tails:XYZ} in vertical displacement and velocity. The curves are contours of $E_z=\Phi(R,z)+\tfrac12v_z^2$ at fixed $R=6.3\,\kpc$, the median radius of the progenitor orbit.}
    \label{fig:tails:z:vz}
\end{figure}

Since tidal ribbons trace all vertical displacements and velocities at each $(R,\phi)$ position within the disc, the detection of one such tidal feature puts strong constraints on the vertical structure of the Galactic mass distribution within its vertical extent at each traced $(R,\phi)$. Fig.~\ref{fig:tails:z:vz} demonstrates that the debris accurately traces surfaces of constant vertical energy $E_z=\tfrac12v_z^2+\Phi(R,z)$, and thus allows to read the run of $\Phi$ with $z$ off such a plot.

Whether such tidal ribbons remain coherent over many Gyrs within the environment of the Galactic disc, where spiral arms and the bar induce significant perturbations, is less clear. In fact, the dissolution of a zero-dimensional progenitor into a two-dimensional ribbon, rather than a one-dimensional stream as in the Galactic halo, is the first step of the accelerated mixing within the Galactic disc.

\section*{acknowledgements}
We thank Ralph Sch\"onrich for many insightful discussions, the reviewer, Jason Sanders, for a prompt and valuable report, Christopher Nixon for suggesting the term `ribbon', and Scott Tremaine as well as Kathryn Johnston for helpful discussions, encouragement, and useful feedback on an early draft. Research in Theoretical Astrophysics at Leicester is supported by STFC grant ST/M503605/1.


\bibliographystyle{mnras}
\bibliography{reference,refs} 

\appendix
\section{\boldmath The Hessian of $H(J)$ for circular orbits in spherical galaxies}
\label{app:A}
For spherical galaxies, the radial motion at fixed angular momentum $L$ is governed by the Hamiltonian $H(r,p_r)=\tfrac12 p_r^2 + \sub{\Phi}{eff}(r)$ with the effective potential
\begin{align}
	\sub{\Phi}{eff}(r)&\equiv \Phi(r)+L^2/2r^2.
\end{align}
The minimum of $\sub{\Phi}{eff}(r)$ occurs at the radius $\sub{r}{c}(L)$ of the circular orbit, where $\sub{\Phi}{eff}(\sub{r}{c}(L))=\sub{E}{c}(L)$, the energy of that orbit. Taylor expanding the effective potential in $x\equiv r-\sub{r}{c}(L)$ gives
\begin{subequations}
\begin{align}
	\label{eq:Phi:eff:Taylor}
	\sub{\Phi}{eff}(r) &= \sub{E}{c}(L) + \tfrac{1}{2!}\kappa^2x^2
	+ \tfrac{1}{3!}\alpha x^3
	+ \tfrac{1}{4!}\gamma x^4 + \mathcal{O}(x^5)
\end{align}
with
\begin{align}
	\kappa^2&= \Phi''(r) + \frac{3L^2}{r^4},\\
	\alpha  &= \Phi'''(r) - \frac{12L^2}{r^5}&
			&= \diff{\kappa^2}{r}-\frac{3\kappa^2}{r},\\
	\gamma 	&= \Phi''''(r) + \frac{60L^2}{r^6}&
			&= \tdiff{\kappa^2}{r} - \frac{3}{r} \diff{\kappa^2}{r} 
				+ \frac{15\kappa^2}{r^2},
\end{align}
\end{subequations}
which are evaluated at $r=\sub{r}{c}(L)$, and hence functions of $L$. Truncating the series~\eqref{eq:Phi:eff:Taylor} at the second order obtains classical epicycle theory \citep{Lindblad1926} with solution $x=\sqrt{2J_r/\kappa}\,\sin\theta_r$ and Hamiltonian $H_0=\sub{E}{c}+\kappa J_r$, the Taylor expansion of $H(J_r,L)$ to linear order in $J_r$. For calculating $\partial^2H/\partial J_r^2$ at $J_r=0$ we require the expansion to order $J_r^2$, which we derive via second-order canonical perturbation theory. Following \citeauthor{LichtenbergLieberman1983} (\citeyear{LichtenbergLieberman1983}, \S2.5),
\begin{align}
	\label{eq:H1}
	H_1&= \frac{\alpha}{3!} x^3
    = \frac{\alpha}{3!} \left[\frac{2J_r}{\kappa}\right]^{3/2}\left(3\sin\theta_r-\sin3\theta_r\right),\\
	H_2&= \frac{\gamma}{4!} x^4
    = \frac{\gamma}{4!} \left[\frac{2J_r}{\kappa}\right]^{2}\left(3-4\cos2\theta_r+\cos4\theta_r\right).
\end{align}
At first order, the correction to the Hamiltonian $\overline{H}_1=\langle H_1\rangle=0$, where $\langle.\rangle$ denotes a phase average, and the generating function
\begin{equation}
	\label{eq:w1}
	w_1 =
	\frac{1}{\kappa}\int(\langle H_1\rangle - H_1)\,\d\theta_r =
		\frac{\alpha}{3!\kappa} \left[\frac{2J_r}{\kappa}\right]^{3/2}
		\left(\cos3\theta_r-9\cos\theta_r\right).
\end{equation}
At second order, the correction to the Hamiltonian
\begin{equation}
	\label{eq:H2}
	\overline{H}_2 = \left\langle H_2 + \tfrac12\big\{w_1,H_1-\langle H_1\rangle\big\}\right\rangle,
\end{equation}
where $\{,\}$ denotes the Poisson bracket. Inserting equations (\ref{eq:H1}-\ref{eq:w1}) into \eqref{eq:H2} and combining with the lower orders, yields
\begin{equation}
	H(J_r,L) = \sub{E}{c}(L) + \kappa(L) J_r + \frac{1}{16}\left[\frac{\gamma}{\kappa^2}-\frac{5}{3}\frac{\alpha^2}{\kappa^4}\right]_{\sub{r}{c}(L)} J_r^2 + \mathcal{O}(J_r^3).
\end{equation}
From this we obtain the Hessian at $J_r=0$ as
\begin{subequations}
\begin{align}
	\ptdiffx{H}{L} &= \diff{\omega}{L}
		= \frac{1}{r^2} \left[1-\frac{4\omega^2}{\kappa^2}\right],
	\\
	\ptdiff{H}{J_r}{L} &= \diff{\kappa}{L}
		= \diff{\kappa^2}{r} \frac{\omega}{r\kappa^3},
	\\
	\ptdiffx{H}{J_r} &= \frac18\left[\frac{\gamma}{\kappa^2}-\frac{5}{3}\frac{\alpha^2}{\kappa^4}\right]
\end{align}
evaluated at $r=\sub{r}{c}(L)$.
\end{subequations}

\bsp
\label{lastpage}
\end{document}